# Emotion Based Prediction in the Context of Optimized Trajectory Planning for Immersive Learning

**Akey Sungheetha**
Associate Professor,
Department of Computer Science and Engineering,
Alliance College of Engineering and Design, Alliance University,
Bangalore, India.
akey.sungheetha@alliance.edu.in,

**Rajesh Sharma R**
Associate Professor,
Department of Computer Science and Engineering,
Alliance College of Engineering and Design, Alliance University,
Bangalore, India.
rajeshsharmar.r@alliance.edu.in

**Chinnaiyan R**
Professor,
Department of Computer Science and Engineering,
Alliance College of Engineering and Design, Alliance University,
Bangalore, India.
chinnaiayan.r@alliance.edu.in

*Abstract*: The use of Google Expedition and touch screen-based emotion in the virtual components of immersive learning is investigated. The objective is to investigate possible ways to combine these technologies to enhance virtual learning environments and learners' emotional engagement. Pedagogical application, affordances, and cognitive load are the corresponding measures that are involved. Students will gain insight into the reason behind their significantly higher post-assessment Prediction Systems scores compared to preassessment scores through this work that leverages technology. This suggests that it is effective to include emotional elements in immersive learning scenarios. The results of this study may help develop new strategies by leveraging the features of immersive learning technology (ILT) in educational technologies to improve virtual reality (VR) and augmented reality (AR) experiences. Furthermore, the effectiveness of immersive learning environments can be raised by utilizing magnetic, optical, or hybrid trackers that considerably improve object tracking.

## 1 AN OVERVIEW OF AUGMENTED AND VIRTUAL REALITY IMMERSIVE LEARNING

The emergence of AR and VR technologies has caused a paradigm change in the educational sector. A growing body of literature has emerged, examining the applications, impacts, and challenges associated with the integration in educational settings, as researchers and educators work together to fully realize the transformative potential of these immersive learning environments. To provide a thorough overview of the state of knowledge regarding the improvement of education, this literature review attempts to synthesize important findings and trends. A theoretical framework for comprehending the pedagogical and cognitive foundations of applications is established by the foundational literature on immersive learning environments. Let's say, the- engaging quality of Experience-s matches up with Vygotsky's social-cultural theory. This theory highlights the- value of learning togethe-r and interacting socially. Research indicates that these technologies have the capacity to generate situated learning opportunities, thereby submerging students in real-world scenarios that promote the transfer and retention of knowledge [1]. A substantial amount of research highlights the beneficial effects on student motivation and engagement. When created well, virtual environments have been demonstrated to hold students' interest and encourage intrinsic motivation [2]. Students can feel more present and fully engaged in their learning experiences because of the novelty and interactivity. Three strategies that have been found to be particularly effective at engaging students in a variety of subjects are virtual field trips, simulations, and interactive 3D models [3]. Improvements in spatial comprehension, problem-solving abilities, and information retention are highlighted by research on the cognitive advantages of education. Students can manipulate virtual objects and participate in hands-on activities thanks to the interactive nature of these technologies, which encourages experiential learning [4]. Moreover, meta-analyses indicate that increased learning outcomes, especially in science and math, are positively correlated with their use [5].

## 2. STUDY ON IMMERSION LEARNING ENVIRONMENT

Literature also identifies implementation-related challenges in the context of education, despite the significant potential benefits. The need for teacher training, financial concerns, and technical limitations keeps coming up [6]. Careful thought should also be given to motion sickness issues, the learning curve of new technologies, and issues of equity and access [7]. The literature foresees innovative future paths for immersive learning environment research and application. The landscape is about to change due to advancements in haptic feedback, augmented reality glasses, and collaborative virtual spaces. Furthermore, investigating tailored and flexible educational opportunities has potential in meeting a range of learning requirements.

Immersion learning environments have a wealth of literature that highlights how these technologies can revolutionize education. The field is dynamic, with ongoing advancements and challenges that require continued investigation, even though the many benefits highlighted by existing studies are not negligible. This review provides a foundation for comprehending the current state of knowledge and opens new avenues for investigation into the various facets of improving education.

## 3. ENHANCING EDUCATION EXPERIENCE WITH AR VR

Autonomous This plan combines different research me-thods. Using a mix, we strive to fully understand how le-arning in immersive settings affe-cts schooling. Surveys, interviews, and evaluations of educational performance will all be part of the research design. Students and teachers from a range of academic backgrounds and levels will be among the varied sample of participants in the study. To guarantee representation across age groups, educational backgrounds, and degrees of experience with AR and VR technologies, the participants will be chosen through purposive sampling. The perception of virtual presence among students was gauged through a virtual presence survey. For the experiment displayed in Figure 1, the ten items were developed at the comprehension level of an elementary student using a seven point Likert scale [8].

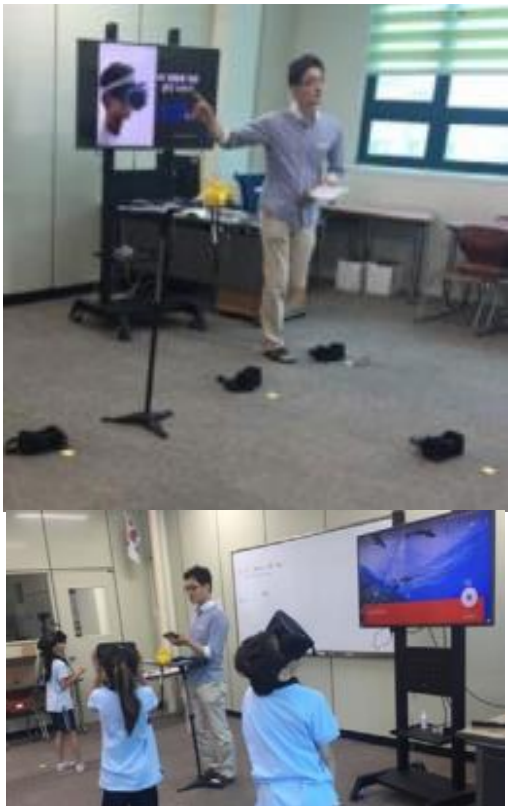

Figure 1. Virtual presence

An explanation is given on how trainee teachers using Google Expedition affe-cted their science-, technology, enginee-ring, and math (STEM) camp teaching during the summer explore. Further, we studied how it impacted stude-nt learning. By conducting semi-structured inte-rviews with a thirteen-question format, our focus was to discover the- trainees' firsthand expe-riences. These-teachers were- using Google Expedition for applying immersive- virtual reality within the framework of a STEM summe-r camp for younger students. We utilize-d straightforward, open-ended que-stions that the interviewe-es could easily understand and e-xpand upon. The central areas of our que-stions were cognitive load, be-nefits, and teaching applications. To enable participants to share their opinions on how AR and VR have affected their teaching or learning experiences, open-ended questions were included in the survey [9].

## 5. EXAMINING EDUCATIONAL OUTCOME ON IMMERSIVE LEARNING INTERVENTIONS

Before studying about poles, the Prediction Systems group's test scores were- compared with the Automatic Control group's scores. The latter had already learned about poles, in a mix of lab and regular classroom settings. Prediction Systems students' post-assessment scores are significantly greater than Automatic Control students' pre-assessment scores, according to the findings of the two-tailed in independent t-tests. The findings indicate that the overall result ($t(74) = 2.99$, $p = 0.0066 < 0.05$) is supported, as are the results for damp ($t(74) = 5.2563$, $p = 1.75 \times 10^{-6} < 0.05$), stabilization ($t(73) = 5.6465$, $p = 3.89 \times 10^{-7} < 0.05$), and polar ($t(77) = 7.0181$, $p = 1.61 \times 10^{-9} < 0.05$). Assessing participants' cognitive abilities, problem-solving abilities, and subject-specific knowledge before and after exposure is a topic of discussion [10]. The following results are obtained in specific educational settings when technology is used to implement immersive learning interventions. This section presents findings from the learner (population), ILT (action), and instruction in architecture (comparison) components built on the PICO concept. The ILT that was used in architecture education has the following features, which also describe how it was put into practice. Study level attained. Various study levels have seen the application of ILT: Year 1 saw 11% in 2 articles, Year 2 saw 25% in 5 articles, Year 3 saw 21% in 4 pieces, and Years 4 and 5 saw 11%. Three pieces. Eleven percent of the publications did not indicate the study level, while another twenty-one percent were applied in four study years. Sort of ILT, really. Four primary forms of ILT were found to have been used in architectural education: (1) five studies on the use of virtual reality (VR); (2) five works on game innovation; (3) four studies on virtual reality environments (VREs); and (4) three studies on augmented reality. AR, VR, and other ILTs have been combined with gaming technology on many occasions. In a single study, the research demonstrated how to integrate two ILT (VRE AR). Create scenarios that support learning objectives and are in line with the curriculum [11]. The following should be kept an eye on both during and after the interventions: participant interactions, engagement levels, and learning outcomes. The intervention sought to address the three constructs—instructor-centric pattern, sparse student-student interactions, and low student participation in information exchange/negotiation—that were identified in the monitoring stage of the online collaborative learning process. The earlier recommendations from Social Network Analysis (SNA) research, which placed an emphasis on group cohesion, interactivity, position in exchange of information, and role in the collaborative group, served as the impetus for the intervention. These recommendations were meant to monitor and direct communicational actions through a more effective collaboration [12]. The following coding strategies were applied to group the answers and derive important information about the participants' experiences in the education section. The 80 papers that were obtained from the second manual filtering are represented as the identified articles in Figure 2. The concept matrix, which demonstrates it as a rational process that defines multiple concepts (such as variables, theories, topics, or methods), was modified by us and used as a classification scheme to put all pertinent articles in one group [13]. We created a preliminary concept matrix based on previously published research, and we added new concepts to it as we classified the data [14].

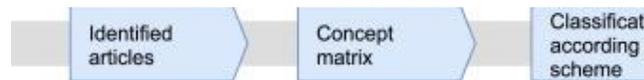

Figure 2. Classification stages

As demonstrated in Figures 3, 4, and 5, the statistically significant differences in educational outcomes linking the control and investigational groups were found through the analysis of random surveys based on t-tests or Analysis of Variance (ANOVA).

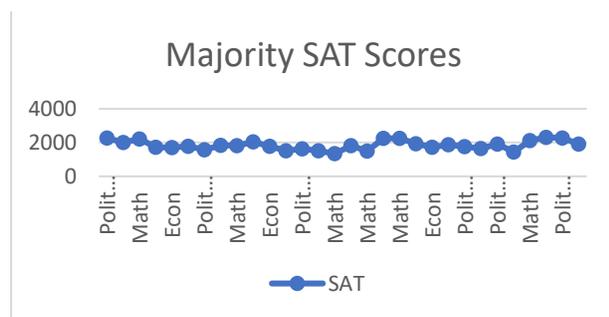

Figure 3. Scholastic Aptitude Test (SAT)

identically informed consent from each participant, guaranteeing the protection of their privacy and confidentiality [16]. Respecting ethical standards when conducting research with human subjects and considering how immersive experiences could affect subjects' feelings and perceptions. The task involved assessing a viewer's emotional response to a video. Because some emotive interactions are intense and continuous self-rating may not be possible, the participants viewed the video twice and rated how they felt each time.

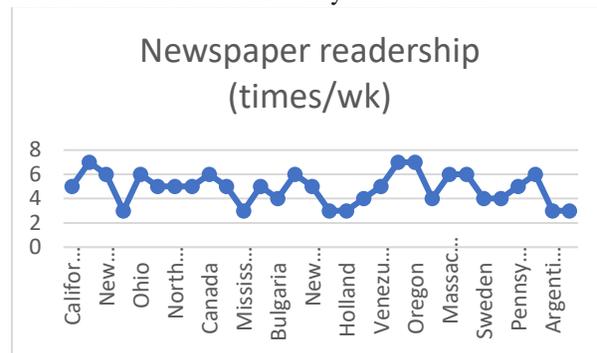

Figure 4. Newspaper readership (times/wk)

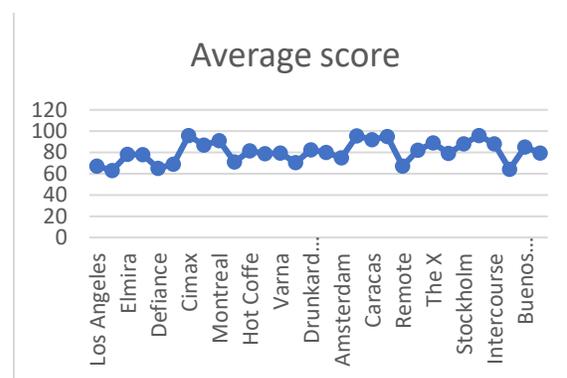

Figure 5. Average score

Using thematic analysis, researchers can find recurrent themes and patterns in the qualitative data collected from interviews. Three primary themes emerge from the results of focused groups (FG) and individual interviews: contextual transfer, obstacles and dangers to actualization, and the inventive, compelling, and powerful nature of virtual reality (VR) [15]. With pediatric- focused articles coming from a single institution, nearly all articles (94percent) detailed efforts in adult populations. Approximately 33% of the articles talked about biobanking programs, and 52% talked about clinical trial consent. The more closely the two ratings matched when sliders were used, the more reliable the sliding models were compared to a touch a screen-based emotion assessment system [17]. 167 participants have undergone the scale's validity and reliability analyses. A structure comprising three factors and fifteen items has been established following the application of necessary steps in EFA. The first factor has been identified as" the use satisfaction," the second as" the use anxiety," and the third as "the use willingness" based on expert opinions and literature. Students' satisfaction levels with augmented reality applications can be found in the first factor, which contains seven positive statements. Six negative statements in the second factor will demonstrate how anxious students are about using augmented reality apps. Lastly, there are two affirmative statements in the third factor that will highlight students' aspirations to use augmented reality apps in the future. Even though the factor's formation from two items may be viewed as a research limitation, the factor's reliability is generally high. Applying Cronbach's alpha to the factors obtained and the entire scale, internal consistency analysis revealed the scale's reliability (the whole scale's α=.835, the first factor's α=.862, the second factor's α=.828, and the third factor's α=.644) [18].

## 6. COMPARATIVE EVALUATION BASED ON OBJECT TRACKING

To ensure the feasibility and efficacy of the research methods, a pilot study involving the following setup is necessary to refine survey instruments, interview protocols, and intervention scenarios. Using a variety of trackers magnetic, optical, mixed, markers, etc. With AR, computer-generated virtual objects may be coordinated by referenced to the world as seen through a camera. The identical kind of marker that was utilized in lab 1 is also used in lab 2. The tracking method is obviously very cheap (just print the marker), but the algorithm takes a very long time because it must locate the marker, analyze each received video frame, and calculate its position in the space that the camera observes while taking the camera's distortion and intrinsic parameters into account. Due to its requirement for complete visibility in the video image, the marker is susceptible to some constraints. As a result, consistent lighting becomes crucial. Object tracking is significantly improved using optical, magnetic, or hybrid trackers. It frees up the host CPU from tracking-calculus tasks that are typically performed by expensive but extremely accurate outboard processors [19]. With an extensive mixed-methods approach, this methodology seeks to offer a nuanced comprehension of the role that immersive learning environments play in improving education.

## 7. CONCLUSION AND FUTURE DIRECTIONS

With the educational landscape always changing, the objective of this findings was to look into the possibilities for transformation in immersive learning environments. The integration of both qualitative and quantitative data has provided valuable insights into the ways in which augmented and virtual reality (AR) and virtual reality (VR) might enhance the educational experience for both educators and learners. After being exposed to immersive learning environments, students' motivation, engagement, and academic performance increased statistically significantly, according to the quantitative analysis. Participant responses on a Likert scale revealed a discernible change in perspective, with most expressing increased interest and enthusiasm for the interventions that were presented. A positive correlation between academic outcomes and immersive experiences is suggested by the experimental group's notable improvements in cognitive skills and problem-solving abilities. Participant interviews yielded qualitative insights that enhanced the quantitative findings by illuminating the subtleties. A sense of presence, enhanced interactivity, and the possibility of bridging theoretical ideas with practical applications were among the themes that emerged from these talks. Teachers have also emphasized how these technologies can be used in a variety of subject areas and can promote collaborative learning environments. Nevertheless, difficulties and factors also came to light, highlighting the necessity of continued study and advancement in this area. Implementing immersive learning environments successfully is influenced by several factors, including accessibility issues, technical limitations, and the need for extensive teacher training programs. The emotional impact of virtual experiences was also recognized from an ethical perspective, highlighting the significance of thoughtful and responsible integration. By providing dynamic, captivating, and customized learning experiences, AR and VR technologies have the potential to completely transform education, as this exploration has shown. The study's conclusions add to the expanding corpus of research on immersive learning environments by illuminating the ways in which these tools can improve instruction in a variety of contexts. As this investigation comes to an end, AR and VR technologies have the power to completely transform education by providing dynamic, captivating, and customized learning opportunities. The study's conclusions add to the expanding corpus of research on immersive learning environments by illuminating the ways in which these tools can improve instruction in a range of contexts.